\documentclass[english,twocolumn,10pt,api,prb]{revtex4-1}
\usepackage{graphicx}
\usepackage{dcolumn}
\usepackage{bm}
\usepackage{amssymb}
\usepackage{subfigure}
\usepackage{color}
\begin{document}
\title{Influence of Quantum and Thermal Noise on Spin-Torque-Driven Magnetization Switching}

\author{Yong Wang}
\author{Yan Zhou}
\author{Fu-Chun Zhang}
\affiliation{Department of Physics, The University of Hong Kong, Hong Kong SAR, China}

\begin{abstract}
We apply a recently developed quantum theory of spin transfer torque to study the effect of the
quantum noise in spin transfer process on the magnetization switching in spin-torque-driven devices. The quantum noise induces considerable fluctuation of the switching time at zero temperature. By including the thermal noise, the temperature dependence of the expectation value and standard deviation of the switching time are obtained in the Monte Carlo simulations, and the results are fitted to an effective first passage model. We expect that the predictions here are observable in the single-shot measurement experiments.
\end{abstract}
\maketitle

In recent years, spin-transfer torque magnetoresistive random-access memory (STT-MRAM) has attracted extensive studies because of its prospective technology applications.\cite{Review} The basic principle of STT-MRAM is the spin-polarized current driven magnetization switching due to spin transfer in the submicro-sized magnetic structures.\cite{STT1,STT2} Although the magnetization dynamics is described well by the Landau-Lifshitz-Gilbert (LLG) equation for the ideal cases,\cite{Macro} the practical performance of the devices is affected or even is dominated by the magnetization noises in the real systems.\cite{Brown,Exp1,Exp2,Exp3,Exp4,XBWang} For example, in the regime of thermally activated magnetization switching, thermal noise plays the major role in the switching and STT significantly modifies the effective temperature of the devices.\cite{Therm1,Therm2} While in the dynamical regime, the single-shot measurement experiments have demonstrated the stochastic trajectories of the STT-driven magnetization switching caused by the magnetization fluctuations.\cite{SingShot} Thus, understanding the effect of noise on the magnetization switching is desirable for the development of the STT-MRAM devices. In addition to the intrinsic quantum noise of the magnetization, there are three sources of the noises during the spin transfer process : thermal noise from the surrounding environment at finite temperature;\cite{Brown} magnetization noise transferred from the current noise;\cite{FBTB,CSK} quantum noise from the interaction between current and magnetization.\cite{W-S1,W-S2} In this paper, we study the effects of these noises on the magnetization switching dynamics with the quantum theory of STT developed recently.\cite{W-S1,W-S2}

\begin{figure}[ht]
\includegraphics[scale=0.36,clip]{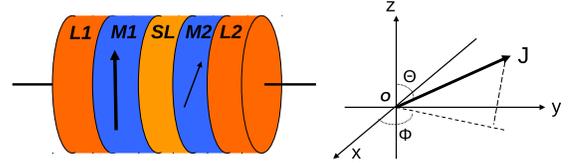}
\caption{(Color online) (Left) Schematic diagram of a spin-transfer torque device.  L1/L2 : metallic contacts;
M1 : fixed FM layer as spin polarizer. M2 : free FM layer. SL : spacer layer.
(right) The coordinate system for a spin coherent state  $|J,\Theta,\Phi\rangle$. Here, x-axis is along the electron transport direction,
and z-axis is along the electron spin-polarizaton direction. }\label{structure}
\end{figure}

The device structure considered here is illustrated in Fig.~\ref{structure}. Between two metallic contacts L1/L2, the pinned polarizer layer M1 and the free ferromagnetism (FM) layer M2 are separated by a non-magnetic spacer layer (SL). For simplicity, we assume that the layer M2 is a single-domain magnetic moment. When the electron current passes through M1 under external bias, it will get spin-polarized and exert spin torque on M2. When the current is larger than a critical value, the magnetization of M2 will be driven out of equilibrium. To include the quantum noise during the spin transfer, we exploit the method developed in Ref.~\onlinecite{W-S1} to treat the STT effect. Here, the magnet of M2 is described by the spin coherent state, and the current acting on M2 will be modelled as a sequence of electrons injected one by one with a fixed time interval $\tau$.\cite{W-S1} Then the stochastic dynamical equation of the spin coherent state $|J,\Theta,\Phi\rangle$ is formally given as
\begin{eqnarray}
i\frac{\partial}{\partial t}|J,\Theta,\Phi\rangle=(\omega_{0}\hat{\mathbf{n}}_{0}+\omega_{T}\hat{\mathbf{n}}_{T})\cdot\hat{\mathbf{J}}|J,\Theta,\Phi\rangle.\label{DynEq}
\end{eqnarray}
Here, $\hat{\mathbf{J}}$ is the macrospin operator; $\omega_{0}\hat{\mathbf{n}}_{0}$ includes the effects of the effective field, Gilbert damping, and possible thermal fluctuation field on the magnet. $\omega_{T}\hat{\mathbf{n}}_{T}$ gives the quantum spin torque, which naturally takes account into the quantum noise during spin transfer.\cite{W-S1}

In the simulations, we set the magnetization of M2 as $M$=1.27$\times10^{6}$~A/m, the cross-section area as $A=8100$~nm$^{2}$, and the thickness as $d=1$~nm. Then the spin quantum number $J$ of M2 is estimated as $J=\frac{MAd}{\gamma\hbar}$, where $\gamma$ is the electron gyromagnetic ratio, and $\hbar$ is the reduced Planck constant. The fields $\omega_{0}\hat{\mathbf{n}}_{0}$ in Eq.~(\ref{DynEq}) are dependent on the configuration of the simulated device.  We define the direction of electron current along the $x$-direction, and the electron spin polarization along $z$-direction, as shown Fig.~\ref{structure}. The thin film structure of M2 causes a strong demagnetization field -$H_d  m_x \hat{e}_x$ in $x$-axis, and the anisotropy of film shape causes an uniaxial anisotropy field $H_{k} m_z\hat{e}_z$ along $z$-axis, where $m_x$ and $m_z$ are the x- and z- components of the unit vector of the free layer magnetization, respectively. We set $H_{d}=4\pi M=1.6$~T and $H_{k}=0.03$~T as in the former studies with the classical method.\cite{macrospin} Besides, the Gilbert damping coefficient is set as $\alpha=0.02$. The STT term $\omega_{T}\hat{\mathbf{n}}_{T}$ is included in the same way as Ref.~\onlinecite{W-S1}. The time interval $\tau$ between two successively injected electrons is estimated from the current $I$, and the spin state of each electron is set as up (+) or down (-) randomly according to the spin-polarization ratio $p$ of the injected current. To calculate the scattering matrix,\cite{W-S1} the electron velocity is set as $v_{g}=3.5\times10^{7}$~cm/s, and the   spin-dependent potentials are set as $V_{\pm}=0.7\pm0.6$~eV. Here, since the inhomogeneity of electron velocity have little effect on the magnetization noise,\cite{W-S1} we take the same value $v_{g}$ for all the electrons.  

\begin{figure}[ht]
\includegraphics[scale=0.3]{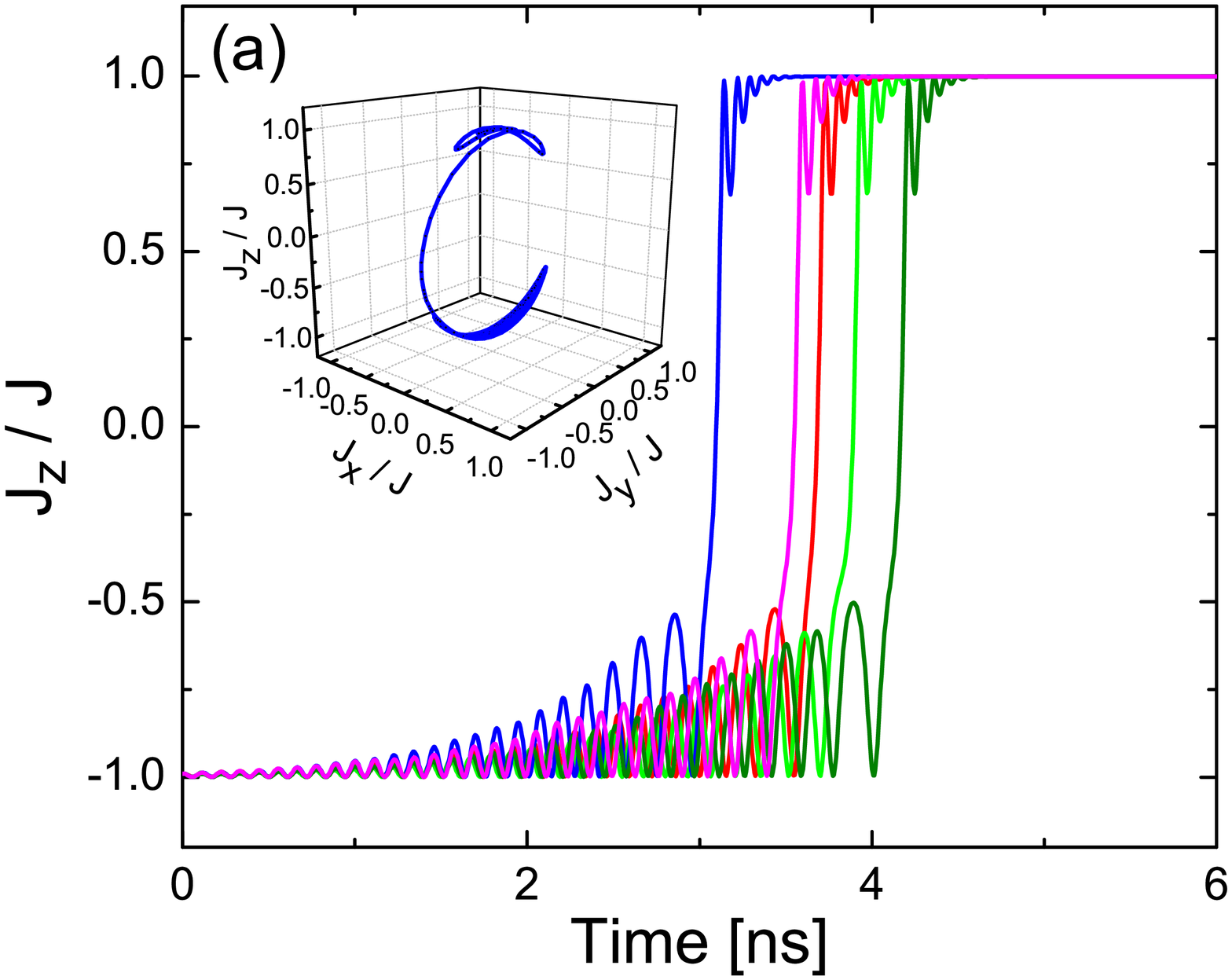}
\includegraphics[scale=0.3]{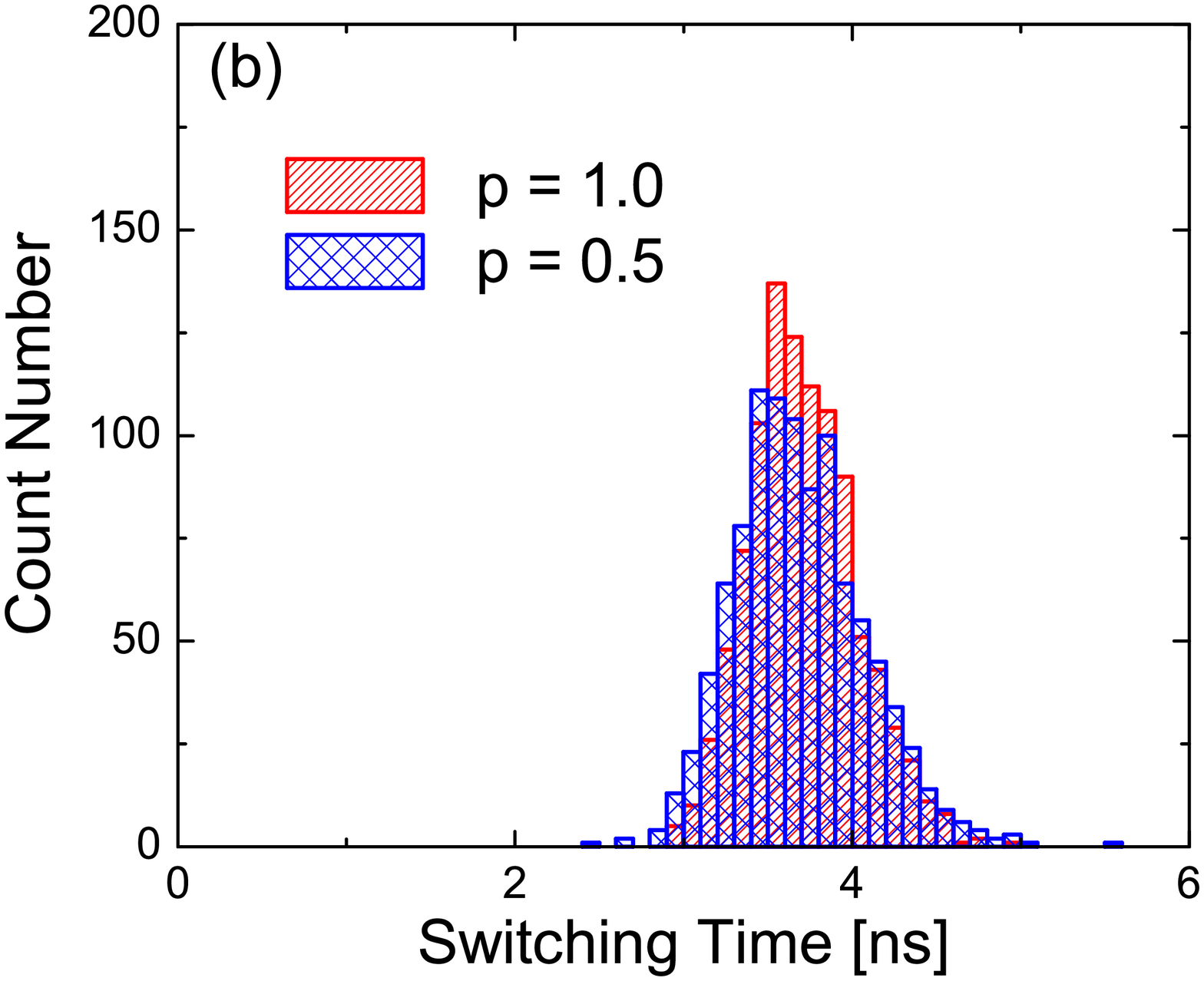}
\caption{(Color online) (a) Five simulated trajectories of the z-component of the normalized magnetization $\mathsf{J}_{z}/J$ at zero temperature. Insert : one sample trajectory of magnetization switching from $\Theta=\pi/2$ to $\Theta=0$. The simulation parameters are given in the body text. (b) Histograms of magnetization swiching times at zero temperature. Red : I=0.9~mA, p=1.0; Blue: I=1.8~mA, p=0.5. For each case, 1000 sample points are exploited. }\label{CountP}
\end{figure}

In Fig.~\ref{CountP} (a), the trajectories of the magnet driven by STT at zero temperature is shown. Here, the current is set as $I=0.9$~mA, and $p=1.0$. The initial position of the magnet is set at $\Theta=3.0,~\Phi=\pi/2$. Due to the strong demagnetization field in $x$-direction, the motion of the magnet is almost restricted in the $y$-$z$ plane. Before the switching happens, the magnet oscillates around the unstable fixed point $\Theta=\pi$. Then the magnet will switch to the region $\Theta < \pi/2$, and approach the stable fixed point $\Theta=0$ spirally. Due to the quantum noise during the spin transfer process,\cite{W-S1,W-S2} the magnetization trajectory will be stochastic in each simulation even with the same parameters. Five trajectories of the z-component of the normalized magnetization $\mathsf{J}_{z}/J$ are demonstrated in Fig.~\ref{CountP} (a).  The switching time $t_{s}$, defined as the time when $\Theta$ is $\pi/2$, varies from trajectory to trajectory. The similar feature was shown in the single-shot experiments of magnetization switching,\cite{SingShot} but the randomness here is completely caused by the quantum noise since there is no thermal noise at the zero temperature. To identify how large the effect of quantum noise can cause, the statistical distribution of switching times for 1000 simulations is shown in Fig.~\ref{CountP} (b) as the red histogram. We found that the quantum noise causes a rather wide distribution of the switching time from 2.8~ns to 4.6~ns around the average value 3.7~ns. This is a rather large effect and is observable by the single-shot measurements.\cite{SingShot} For comparison, we also consider a partially polarized current (I=1.8~mA, p=0.5) which gives the same spin torque but includes additional noise from the current,\cite{W-S1} and the result is also shown in Fig.~\ref{CountP} (b) as the blue histogram. We found that the distribution of switching time becomes a little broader and the average switching time has decreased to 3.68~ns. Thus the fluctuation of the switching time at zero temperature is mainly due to the quantum noise generated in the spin transfer process, and the injected current noise has little effect. This is due to the fact that the magnitudes of the quantum noise are in the same order for the two cases.\cite{W-S1,W-S2}  

\begin{figure}[ht]
\includegraphics[scale=0.30]{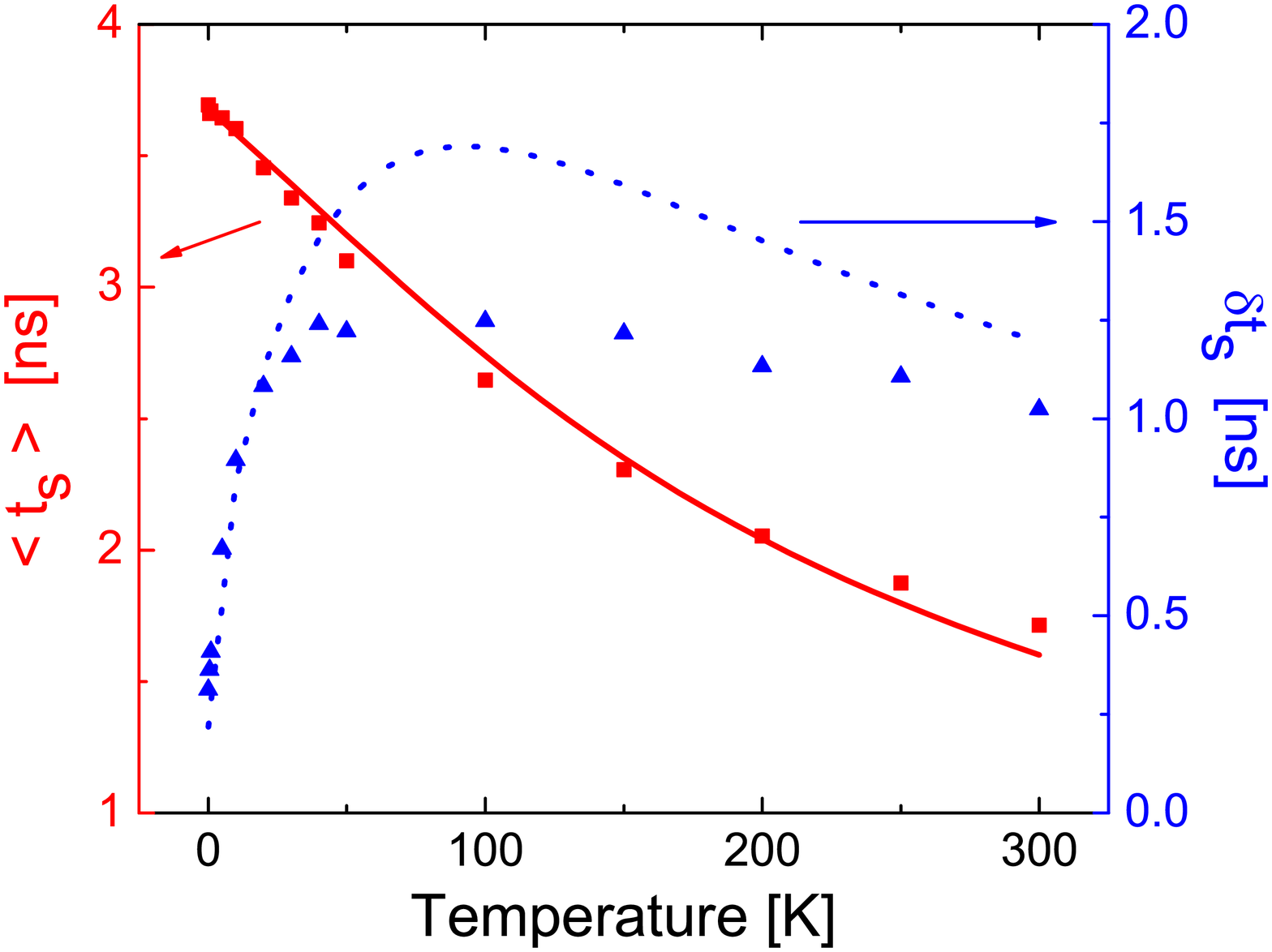}
\caption{(Color online) Dependence of the mean switching time $\langle t_{s}\rangle$ and standard deviation $\delta t_{s}$ on the temperature $T$. Red square : $\langle t_{s}\rangle$ from numerical simulations.
Blue triangle : $\delta t_{s}$ from numerical simulations. 5000 simulated trajectories are exploited for each temperature. Red real line : fitting curve for $\langle t_{s}\rangle$ from Eq.~(\ref{first}).
Blue dotted line : fitting curve for $\delta t_{s}$ from Eq.~(\ref{first}) and Eq.~(\ref{second}).}  \label{Temp}
\end{figure}

The effect of thermal noise on the magnetization switching can be included in Eq.~(\ref{DynEq}) by adding a fluctuating magnetic field $\mathbf{h}_{r}$ defined as\cite{Brown,Therm1,Therm2}
\begin{eqnarray}
\langle h_{r}^{i}(t)\rangle=0,\quad\gamma^{2}\langle h_{r}^{i}(t)h_{r}^{j}(t')\rangle=2D_{T}\delta_{ij}\delta(t-t'),\label{MagF}
\end{eqnarray}
with the thermal noise correlator $D_{T}=\frac{\gamma\alpha k_{B}T}{ MAd}$.  Here, $i$ and $j$ are Cartesian indices, $k_{B}$ is the Boltzmann constant, $T$ is the temperature. For each of the given temperatures $T$=0K, 0.5K, 1K, 5K, 10K, 20K, 30K, 40K, 50K, 100K, 150K, 200K, 250K, 300K, we performed the simulations according to Eq.~(\ref{DynEq}) for 5000 times and got the statistical distribution of the switching time $t_{s}$. Then the dependence of the mean value $\langle t_{s}\rangle$ and the standard deviation $\delta t_{s}$ of the switching times on the temperature $T$ were obtained, as shown in Fig.~\ref{Temp}.  We found that as the temperature increases, the mean switching time $\langle t_{s}\rangle$ decreases monotonically (square symbols), i.e. the thermal noise will assist the magnetization switching for the given STT. On the other hand, the standard deviation of switching time $\delta t_{s}$, starting from a non-zero value at zero temperature due to the quantum noise, will first increase and then decrease as the temperature increased from $T=0$~K to $T=300$~K (triangle symbols). This is contrary to the intuition that larger thermal noise should cause larger uncertainty of the switching time. However, considering that the mean switching time $\langle t_{s}\rangle$ decreases with the temperature, the relative fluctuation of the switching time indeed increases with the temperature. 

In order to understand the above simulation results, we analyze the corresponding Fokker-Planck equation (FPE)\cite{Brown,Therm1,Therm2,CSK,W-S2} for the probability distribution function $P(\Theta,\Phi,t)$, where
\begin{eqnarray}
\frac{\partial}{\partial t}P=-\nabla\cdot(\mathbf{T}P)+\nabla^{2}(\mathcal{D}P)\label{FPE1}
\end{eqnarray}
Here, the drift coefficient $\mathbf{T}$ comes from all the deterministic terms including the STT in Eq.~(\ref{DynEq}), and the diffusion coefficient $\mathcal{D}$ consists of the contributions from the quantum noise and thermal noise. Integrating Eq.~(\ref{FPE1}) over $\Phi$ in the spherical coordinates,\cite{W-S2} Eq.~(\ref{FPE1}) is formally reduced to
\begin{eqnarray}
\frac{\partial}{\partial t}\overline{P}&=&\frac{\partial}{\partial\Theta}(-\overline{T}_{\Theta}\overline{P})+\frac{\partial}{\partial\Theta}[\sin\Theta\frac{\partial(\frac{\mathcal{D}}{\sin\Theta}\overline{P})}{\partial\Theta}].\label{FPE2}
\end{eqnarray}
Here, $\overline{P}(\Theta)=\int_{0}^{2\pi}\sin\Theta Pd\Phi$, with the normalization condition $\int_{0}^{\pi}\overline{P}d\Theta=1$; $\overline{T}_{\Theta}=\frac{1}{\overline{P}}\int_{0}^{2\pi}\sin\Theta T_{\Theta}Pd\Phi$. $\mathcal{D}$ is independent on $\Phi$ here.\cite{W-S2} Thus, the magnetization switching reduces to a first-passage process\cite{FirstPass} for a particle trapped in the interval $\Theta\in[\pi/2, \pi]$, which satisfies Eq.~(\ref{FPE2}) with the reflecting boundary at $\Theta=\pi$ and absorbing boundary at $\Theta=\pi/2$. Because the expression of $\overline{T}_{\Theta}$ is unknown, the exact solution of Eq.~(\ref{FPE2}) is unavailable. As a semi-quantitative estimation, we approximate Eq.~(\ref{FPE2}) by a drift-diffusion equation with constant coefficients, i.e.
\begin{eqnarray}
\frac{\partial}{\partial t}\overline{P}&\approx&-T^{*}\frac{\partial}{\partial\Theta}\overline{P}+\mathcal{D}^{*}\frac{\partial^{2}}{\partial\Theta^{2}}\overline{P}.\label{FPE3}
\end{eqnarray}   
For convenience, we set the coefficients $T^{*}=\beta_{1}\mathcal{A}$ and $\mathcal{D}^{*}=\beta_{2}(\frac{\mathcal{A}}{2J+1}+D_{T})$, where $\mathcal{A}$ is the magnitude of of the Slonczewski spin torque in $T_{\Theta}$, and $\frac{\mathcal{A}}{2J+1}$ is the coefficient of the quantum noise correlator.\cite{W-S2} $\beta_{1}$ and $\beta_{2}$ are two dimensionless parameters to be determined by fitting the simulations results from Eq.~(\ref{DynEq}) to the analytical results from Eq.~(\ref{FPE3}).

Solutions to the first-passage problem for Eq.~(\ref{FPE3}) gives the first two moments of the switching time as\cite{FirstPass}
\begin{eqnarray}
\langle t_{s}\rangle&=&\frac{\pi^{2}}{4D^{*}}(\frac{1}{2P_{e}}-\frac{1-e^{-2P_{e}}}{4P_{e}^{2}}),\label{first}\\
\langle t_{s}^{2}\rangle&=&(\frac{\pi^{2}}{4D^{*}})^{2}[\frac{1}{4P_{e}^{2}}-\frac{2-(6P_{e}+1)e^{-2P_{e}}-e^{-4P_{e}}}{8P_{e}^{4}}]\label{second}.
\end{eqnarray}
Here, the P\'{e}clet number $P_{e}$ is given as $P_{e}=\pi T^{*}/4\mathcal{D}^{*}$. With the simulation parameters used in Eq.~(\ref{DynEq}),  $\mathcal{A}$ is calculated as 3.5~rad/ns. By choosing the fitting parameters $\beta_{1}=0.121, \beta_{2}=350$, we got the temperature dependence of $\langle t_{s}\rangle$ and $\delta t_{s}$, as shown in Fig.~\ref{Temp}. Here, $\beta_{1}\sim\mathcal{O}(10^{-1})$ and $\beta_{2}\sim\mathcal{O}(10^{2})$ imply that the oscillations of the magnet before switching reduces the effective STT but amplifies the uncertainty. We found that such a simplified model gives the excellent fitting curve of $\langle t_{s}\rangle$ for all the temperatures, but the fitting curve of $\delta t_{s}$ is not so good for the temperatures higher than 20~K. This can be attributed to the over simplification from Eq.~(\ref{FPE2}) to Eq.~(\ref{FPE3}). However, the main features of $\delta t_{s}$ are still captured by this simple treatment. We notice that the first passage model was applied to the thermally activated regime for magnetization switching in a recent work,\cite{pinna} while the results here are in the STT-driven dynamical regime and the quantum and thermal noise are perturbative effects. 

In conclusion, we have studied the influence of the quantum and thermal noise on the magnetization switching driven by STT. At zero temperature, the stochastic nature of the spin transfer torque on the typical nanomagnet causes significant fluctuation of the switching time in the order of nanosecond. As the temperature increases and the thermal noise plays the dominant role, the mean switching time decreases monotonically, and the standard deviation of the switching time first increases and then decreases. These results agree reasonably with a simple first-passage model, and are expected to be observed by the single shot measurements.

This research was supported by the Hong Kong University Grant Council (AoE/P-04/08).

\end{document}